\documentclass[aps,prb,superscriptaddress,preprint,showpacs,floats]{revtex4}
\usepackage{graphicx}
\usepackage{amsmath}
\usepackage{amssymb}
\usepackage{bm}
\usepackage{mathtools}
\usepackage[x11names,svgnames]{xcolor}
\begin{document}

\title{Raman Signatures of Strong Kitaev Exchange Correlations in (Na$_{1-x}$Li$_x$)$_2$IrO$_3$ : Experiments and Theory}

\author{Satyendra Nath Gupta$^1$, P. V. Sriluckshmy$^2$, Kavita Mehlawat$^3$, Ashiwini Balodhi$^3$, Dileep K Mishra$^1$, D.V.S.Muthu$^1$, S. R. Hassan$^2$, Yogesh Singh$^3$, T. V. Ramakrishnan$^1$ and A. K. Sood}
\thanks{To whom correspondence should be addressed}
\email{asood@physics.iisc.ernet.in}
\affiliation{Department of Physics, Indian Institute of Science, Bangalore-560012,India
$^2$The Institute of Mathematical Sciences, C.I.T. Campus, Chennai 600 113, India
$^3$Indian Institute of Science Education and Research (IISER) Mohali, Knowledge City, Sector 81, Mohali 140306, India}

\date{\today}

\begin{abstract}
Inelastic light scattering studies on single crystals of (Na$_{1-x}$Li$_x$)$_2$IrO$_3$ ($x = 0, 0.05$ and $0.15$) show a polarization independent broad band at $\sim $~2750 cm$^{-1}$ with a large band-width $\sim 1800$~cm$^{-1}$.  For Na$_2$IrO$_3$ the broad band is seen for temperatures $ \leq 200$~K and persists inside the magnetically ordered state.  For Li doped samples, the intensity of this mode increases, shifts to lower wave-numbers and persists to higher temperatures.  Such a mode has recently been predicted (Knolle et.al.) as a signature of the Kitaev spin liquid.  We assign the observation of the broad band to be a signature of strong Kitaev-exchange correlations.  The fact that the broad band persists even inside the magnetically ordered state suggests that dynamically fluctuating moments survive even below $T_{N}$. This is further supported by our mean field calculations. The Raman response calculated in mean field theory  shows that the broad band predicted for the spin liquid state survives in the magnetically ordered state  near the zigzag-spin liquid phase boundary.  A comparison with the theoretical model gives an estimate of the Kitaev exchange interaction parameter to be $J_K\approx 57$~meV.
\end{abstract}
\pacs{78.30.Am, 75.10.Kt, 75.10.Jm, 78.20.Ls}
\maketitle
Classical aspects of magnetic frustration have been investigated for a long time; eg.  the fact that globally incompatible magnetic interactions due to disorder lead to a glassy state (spin glasses) \cite{mydosh2015} . Quantum consequences of geometrical and anisotropy related frustration have been actively explored in the last decade or more \cite{moessner2006}.  Magnetic frustration can lead to novel ground states like spin-liquid (SL) and spin-ice.  The search for SL states has traditionally relied on antiferromagnetically coupled spins on lattices with triangular motifs.  Anisotropic magnetic interactions can provide a novel alternate route to such SL states.  This activity received impetus from the work of Kitaev who showed that the exact ground state of a certain kind of intrinsically frustrated spin system on a honeycomb lattice is a quantum spin liquid \cite{kitaev2006}.  Layered honeycomb iridates $A_2$IrO$_3$ ($A =$ Na, Li) with stong spin-orbit coupling were proposed to be avenues where Kitaev like interactions might be realized \cite{jackeli2009}.  Additionally, the real materials would have direct- and super-exchange Heisenberg exchange as well.  Such a Kitaev-Heisenberg (K-H) model for $A_2$IrO$_3$ was studied and based on the relative strength of the two terms, three magnetic ground states were found. A simple Neel anti-ferromagnet in the Heisenberg limit, a quantum spin-liquid (QSL) in the Kitaev limit, with an unusual stripy magnetic order in the middle were predicted \cite{chaloupka2010kitaev}.  These predictions have led to a flurry of activity on the honeycomb lattice iridates $A_2$IrO$_3$ ($A =$~Na, Li) \cite{shitade2009quantum,singh2012relevance,choi2012spin,ye2012direct,chaloupka2010kitaev,kitaev2006,singh2010antiferromagnetic,kimchi2011kitaev,chaloupka2013zigzag,PhysRevB.89.241102,gretarsson2013crystal,comin20122,katukuri2014kitaev,rau2014generic,reuther2014spiral,schaffer2012quantum,sizyuk2014importance}.  Inspite of such intense scrutiny, a minimal model for these materials and in particular, how close the real materials are to the SL state in the dominant Kitaev limit is still unclear.

Although a recent ultrafast optical study on Na$_2$IrO$_3$ has claimed to see signatures of such a SL state in the confinement-deconfinement transition of spin and charge excitations across $T_N$ \cite{PhysRevLett.114.017203}, smoking gun evidence of the Kitaev spin-liquid or of dominant Kitaev interactions has been missing.  The first direct evidence of dominant bond-dependent magnetic interactions has been found very recently using diffuse RIXS measurements on Na$_2$IrO$_3$ \cite{chun2015direct}.  Additionally, this study showed that short ranged zig-zag correlations which are present above $T_N = 15$~K do not change as the temperature is lowered into the magnetically ordered state suggesting that dynamically fluctuating moments survive deep into the ordered state \cite{chun2015direct}.  This study therefore suggests that Na$_2$IrO$_3$ may be close to the Kitaev QSL \cite{chun2015direct}. 

A novel prediction has recently been made for observing the signatures of the Kitaev QSL in Raman scattering on  Na$_2$IrO$_3$ in the form of a polarization-independent broad band response centred at $6J_K$ ($J_K$ is the Kitaev interaction strength) \cite{knolle2014}.  In these calculations the Heisenberg interaction ($J_H$) is assumed to be a weak perturbation ($J_H/J_K = 0.1$).  Predictions of a broad polarization independent Raman band seems to be a generic feature of spin-liquids with similar features being predicted\cite{cepas2008, ko2010} and observed\cite{wulferding2010} for Herbertsmithite ZnCu$_3$(OH)$_6$Cl$_2$.  The broad band has been argued to be associated with excitations of a gapless SL ground state \cite{wulferding2010}.  Thus Raman scattering seems to be a new tool to look for signatures of QSL's.  Recent Raman scattering measurements on the honeycomb lattice Ruthenate $\alpha$-RuCl$_3$, another material proposed for realization of K-H physics, has also revealed a broad continuum of excitations which was interpreted as resulting from proximity to the QSL phase in the strong Kitaev limit \cite{sandilands2015scattering}.                          

We present here the first results of a comprehensive Raman study on the honeycomb lattice iridates ($Na_{1-x} Li_x )_2 IrO_3$  ($x = 0, 0.05, 0.15$).  These materials have previously been reported to have zigzag magnetic long range order for $T_N \approx 15$~K, 10~K, and $6$~K, respectively \cite{singh2012relevance, manni2014effect, PhysRevB.88.220414}.  The series (Na$_{1-x}$Li$_x$)$_2$IrO$_3$ presents an interesting possibility to test the predictions of Ref.~[23].  The structure of Na$_2$IrO$_3$ is far from the ideal undistorted structure required for the super-exchange Heisenberg term to exactly cancel and leave the system in the Kitaev dominant limit \cite{jackeli2009, chaloupka2010kitaev}.  However, Li substitution reduces these distortions \cite{manni2014effect} and can be expected to tune the material towards the ideal structure required for dominant Kitaev physics.  This is indeed suggested by the reduction of $T_N$ and an enhanced frustration parameter $f = \theta/T_N$ with increasing Li content $x$~~ \cite{manni2014effect, PhysRevB.88.220414}. This implies that if Raman scattering can probe Kitaev correlations, then these signatures should be stronger for the Li substituted samples.  With this hypothesis in mind, we have studied the Raman response of (Na$_{1-x}$Li$_x$)$_2$IrO$_3$ ($x = 0, 0.05, 0.15$) to examine the predictions made in Ref.~[23].

Our main results are that for Na$_2$IrO$_3$, we observe a polarization independent broadband Raman signal peaked at $2750$~cm$^{-1}$ with full width at half maximum (FWHM)  $\sim 1800$~cm$^{-1}$.  From the peak position we make a first experimental estimate of the Kitaev interaction to be $J_K \approx 57$~meV\@.  The intensity of this signal increases for Li substituted crystals (Na$_{1-x}$Li$_x$)$_2$IrO$_3$ ($x = 0.05, 0.15$) which, as discussed above, is an expected response of enhanced Kitaev correlations.  The existence of such a Raman mode has so far only been theoretically demonstrated for the Kitaev spin liquid ground state \cite{knolle2014}.  We show using a Kitaev-Heisenberg model treated in a generalized mean field theory that even in the magnetically ordered state close to the SL phase boundary, the predicted Raman response is a broad Kitaev SL like spectrum quite similar to what we observe.  We believe that this demonstrates the survival of Kitaev spin correlations in spite of long range magnetic order.  

Some implications of this discovery are that some of the quantum entanglement and coherence manifest in Kitaev spin liquid systems could survive recognizably and usefully in real systems with magnetic long range order with possible applications, and  also that there is hope for finding realistic materials which have Kitaev spin liquid like ground states whose exact solubility lays bare the nature of low lying excitations (Majorana fermions) and quantum coherence.     
                        
Raman experiments were carried out on freshly cleaved surfaces of single crystalline (Na$_{1-x}$Li$_x$)$_2$IrO$_3$ ($x = 0,0. 05, 0.15$) grown by a self-flux method with excess IrO$_2$ as described elsewhere \cite{singh2010antiferromagnetic}. 
The crystals (Na$_{1-x}$Li$_x$)$_2$IrO$_3$ ($x = 0,0. 05, 0.15$) are characterized by powder and single crystal x-ray diffraction, and magnetic susceptibility measurements performed on the VSM option of a PPMS system (M/S Quantum Design).  Raman  measurements at room temperature as well as in the temperature range of 80 to 400~K were performed using LABRAM HR-800 spectrometer equipped with $532$~nm excitation laser.  The low temperature Raman measurements in the temperature range $300$~K to $4$~K were performed in $180 ^o$  geometry using Ar-ion laser with wavelength 514.5 nm (Coherent Innova 300) and Raman spectrometer (DILOR XY) coupled to a liquid nitrogen cooled CCD detector.  Temperature variation was done with continuous flow He cryostat (Oxford Instrument) from 4K to 300K. The temperature accuracy was $\approx \pm 1$~K\@.

The results of our x-ray diffraction and magnetic measurements are consistent with the previous report \cite{manni2014effect}.  Specifically, we found the (Na$_{1-x}$Li$_x$)$_2$IrO$_3$ crystals to crystallize in the reported mono-clinic space group C2/m ($\# 12$).  Various structural and magnetic property parameters are given in Table 1.


\begin{table}[h]
 \resizebox{0.48\textwidth}{!}{ 
  \begin{tabular}{ | l | l | l | l | l | l|l|l|}

    \hline
    \multicolumn{8}{|l|}{Table 1} \\
    \hline
     x &  a(~\AA) & b(~\AA) & c(~\AA) & $\beta(~^o)$ & T$_N$~\rm({K}) & $\theta$~\rm({K}) & f \\ \hline
    0 &   5.427 &  9.395 &  5.614 & 109.04 & 15 & -120 & 8 \\ \hline
    0.05 &  5.401 & 9.345 & 5.612 & 108.91 & 12 & -110 & 8.4 \\ \hline
   0.15  & 5.355 & 9.258 & 5.612 & 108.65 & 8 & -87 & 10.9 \\
    \hline
  \end{tabular}
  }
  \label{Table 1}
\end{table}


The above results demonstrate that increasing amounts of Li is being successfully substituted into the crystals and the magnetic order is suppressed and the frustration parameter $f = |\theta|/T_N$ increases with increasing Li content.

\begin{figure}[h]
\includegraphics[width=\textwidth]{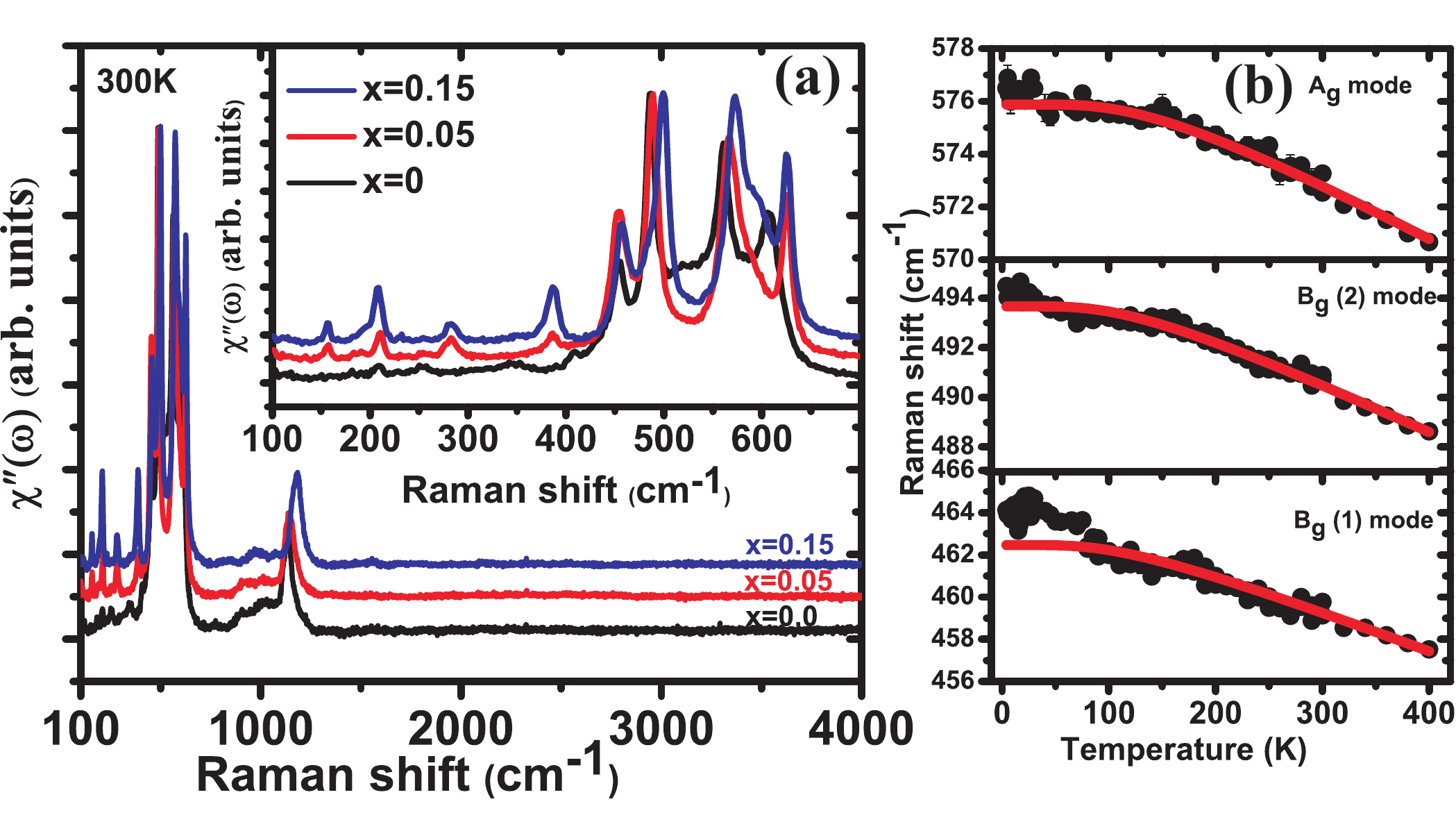}
\caption{(Color online) (a) Raman spectra of (Na$_{1-x}$Li$_x$)$_2$IrO$_3$  single crystals with $x = 0, 0.05, 0.15$ measured at $T = 300$~K in the spectral range 100 to 4000 cm$^{-1}$. Inset: Raman spectra for the spectral range 100 to 700 cm$^{-1}$ to highlight the phonon modes. (b) Temperature dependence of phonon frequencies of the B$_g$(1), B$_g$(2) and A$_g$ Raman modes from 400K to 4K. The solid lines are fit to a simple cubic anharmonic model to the experimental data points.
\label{Fig-Raman-300K}}
\end{figure}

For the monoclinic structure, group theory predicts a total of $36~\Gamma$- point phonon modes (8A$_u$ + 7A$_g$ + 8 B$_g$ + 13 B$_u$); out of which 15 modes (7A$_g$ + 8 B$_g$) are Raman active.  Fig.~\ref{Fig-Raman-300K}(a)  shows the $T = 300$~K Raman susceptibility $\chi^{\prime\prime}$ ($\omega$)  ( $\chi^{\prime\prime}$($\omega$) = Intensity($\omega$)/(n($\omega$)+1), where n($\omega$)+1 is the Bose-Einstein factor) of (Na$_{1-x}$Li$_x$)$_2$IrO$_3$ ($x = 0, 0.05, 0.15$) single crystals in the spectral range 100 to $4000$~cm$^{-1}$.  The region from 100 to 700~cm$^{-1}$ covering seven first order Raman modes is shown in inset of Fig.~\ref{Fig-Raman-300K}(a) and the spectral region between 900 to 1200~cm$^{-1}$ shows second order Raman modes.
The Raman modes at 460~cm$^{-1}$, 490~cm$^{-1}$, 570~cm$^{-1}$ (labelled as B$_g$(1), B$_g$(2) and A$_g$ respectively based on their polarization dependence) are the most prominent Raman modes of the system. The lineshape parameters - mode frequencies and linewidths  extracted from the Lorentzian fit, show normal temperature  dependence (Fig.~\ref{Fig-Raman-300K}(b)). The solid lines are fit to a simple cubic anharmonic model where the phonon decays into two phonons of equal frequency giving a temperature dependence  $\omega(T) = \omega(0)+C[1+2n(\omega(0)/2)]$ , where $n(\omega) = 1/(exp(\hbar\omega/k_BT)-1)$ is the Bose-Einstein mean occupation number and C is the self energy parameter.  All the three Raman modes show normal temperature  dependence of mode frequencies and linewidth (data not shown) as expected due to cubic anharmonic interaction with a small deviation at low temperature in frequency (see lower panel in Fig.~\ref{Fig-Raman-300K}(b)) which can be due to presence of spin phonon coupling.

\begin{figure}[t]
\includegraphics[width=\textwidth]{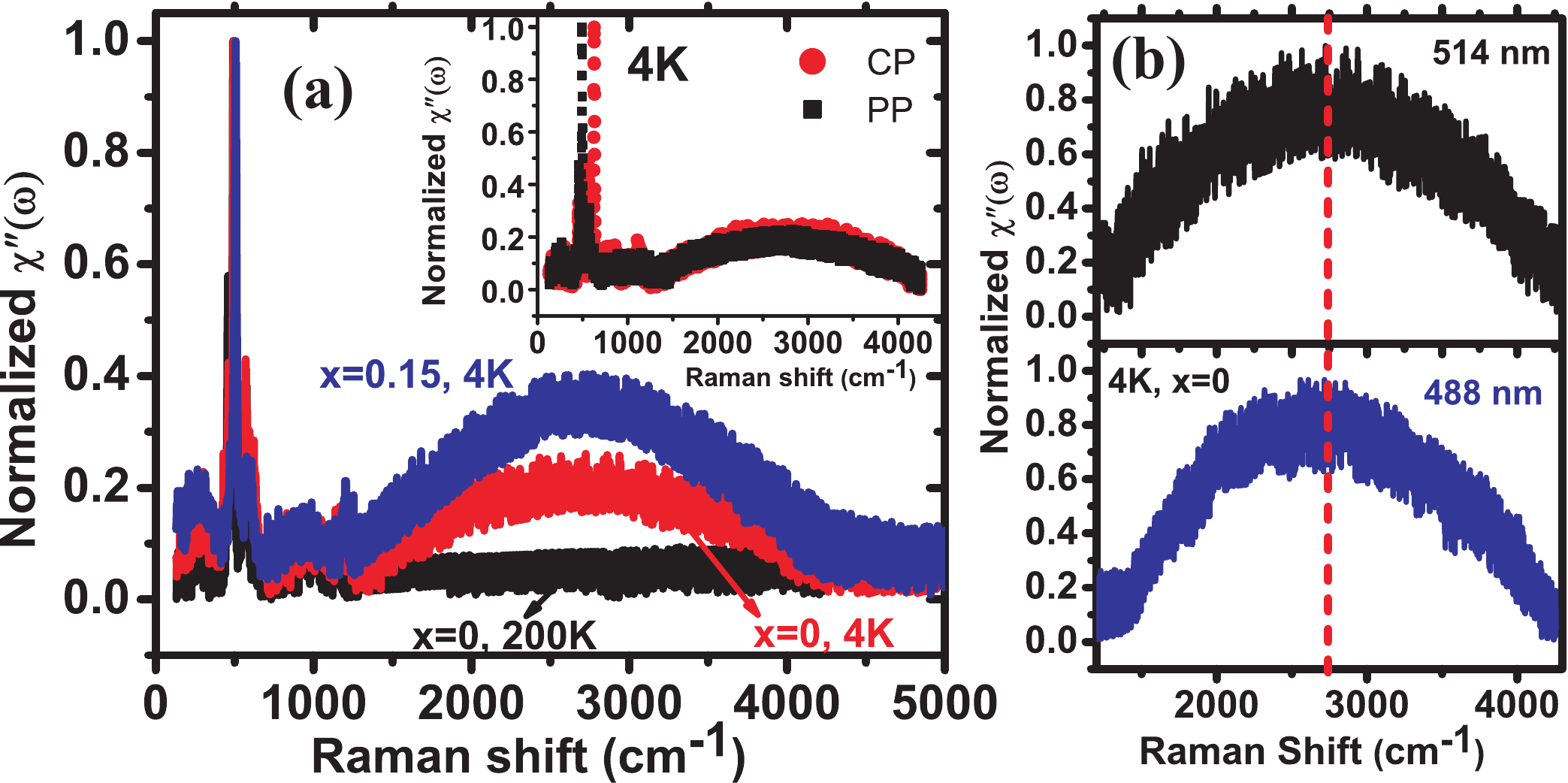}

\caption{ (a) Raman spectra of (Na$_{1-x}$Li$_x$)$_2$IrO$_3$  single crystals at  4~K (for x=0, 0.15) and  200~K (for x=0). The Raman susceptibility of the $490$~cm$^{-1}$ phonon mode has been normalized to $1$. Inset: Polarization dependence of the Raman spectra of Na$_2$IrO$_3$ at 4~K\@. The phonon response is different in CP and PP configurations. (b) BRB with two different laser lines 514 nm and 488 nm. The dashed line shows that BRB does not shift with change in lase wavelength.    
\label{Fig-BRB-Na2IrO3}}
\end{figure}

\begin{figure}[t]
\includegraphics[width=\textwidth]{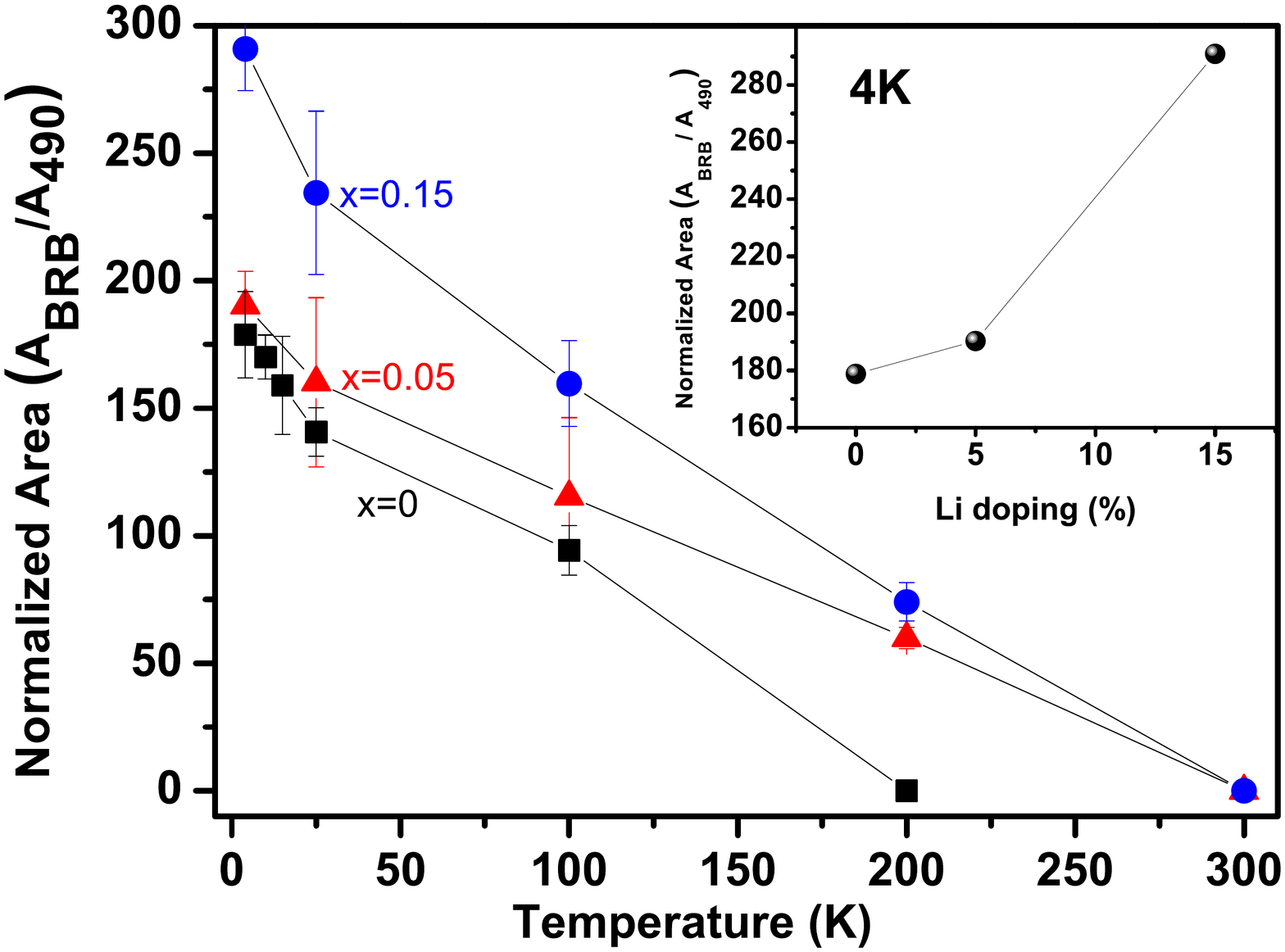}
\caption{Temperature dependence of the broad Raman band centered at $\sim 2750$~cm$^{-1}$.  The intensity decreases with temperature and becomes zero at $\sim 200$~K for Na$_2$IrO$_3$ and at $300$~K for the Li substituted samples. Inset: Area of BRB (normalized with the area of the 490 cm$^{-1}$ phonon mode) as a function of Li doping.     
\label{Fig-BRB-T dependence}}
\end{figure}

Fig.~\ref{Fig-BRB-Na2IrO3} (a) displays Raman susceptibility, $\chi^{\prime\prime}$($\omega$) of (Na$_{1-x}$Li$_x$)$_2$IrO$_3$ single crystals at 4~K (for $x = 0, 0.15$) and at 200~K (for $x = 0$).  The Raman susceptibility corresponding to the phonon mode at $490$~cm$^{-1}$, is only slightly sample and temperature dependent and hence has been normalized to $1$.  The $T = 4$~K spectra show a broad band centred at $\sim 2750$~cm$^{-1}$ which is absent in the high temperature ($T > 200 $~K) data.  In order to rule out luminescence as the origin of the broad band, Raman spectra were recorded with a different laser line (488 nm) at $T = 4$~K and show the same mode without any frequency shift (see Fig.~\ref{Fig-BRB-Na2IrO3} (b)).  This rules out the broad band being related to photoluminescence.  We will henceforth abbreviate the broad Raman band as BRB.  Another important feature of the BRB is its large  band-width of $\approx 1800$~cm$^{-1}$.  We have also looked at possible polarization dependence of this BRB by measurements on Na$_2$IrO$_3$ crystal for incident and scattered polarization in parallel and perpendicular directions and found that they are identical.  This is shown in inset of Fig.~\ref{Fig-BRB-Na2IrO3}~(a) where data measured at $T = 4$~K for cross-polarization (CP) and parallel-polarization (PP) are plotted on top of each other.  No polarization dependence was observed.  The BRB also persists to high temperatures of up to $T = 200$~K for Na$_2$IrO$_3$ (see Fig.~~\ref{Fig-BRB-T dependence}).  

There could be several origins of the BRB.  Since we observe the BRB inside the magnetically ordered phase it could be the two-magnon peak.  In Heisenberg antiferromagnets, the two-magnon Raman mode, seen in systems such as YBa$_2$Cu$_3$O$_6$ \cite{knoll1990temperature} and Sr$_2$IrO$_4$ \cite{cetin2012crossover}, occurs at E$_{2magnon} \sim J_{H}(2sz-1)$, where $s$ is the effective spin of the system and $z$ is the number of nearest neighbors.  Taking $s = 1/2$ and $z = 5$ for a 3-dimensional honeycomb lattice, we expect $E_{2magnon} \sim 4J_{H}$.  Taking a typical estimate for Na$_2$IrO$_3$, J$_H \sim$ 5-12 meV \cite{chaloupka2013zigzag, choi2012spin}, the two magnon Raman band should occur at $\sim 160-380$~cm$^{-1}$.  Our BRB is located at $\sim 2750$~cm$^{-1}$.  We also note that the BRB is observed upto much higher temperatures compared to $T_N$ and therefore cannot be the two-magnon peak.  The BRB could also arise from inter-orbital transitions between $t_{2g}$ and $e_g$ orbitals.  However, energy scales for these transitions are expected to be much larger (in the few~eV range) \cite{gretarsson2013crystal, comin20122}. Another possibility is the transition from the j$_{eff}$ = 3/2 manifold to j$_{eff}$ = 1/2 manifold as possible origin of the BRB.  The experimental fact that the BRB intensity is zero in pure Na$_2$IrO$_3$ above 200K can not be reconciled with this possibility \cite{igarashi2015collective}.
 
This leads us to consider the possibility that the BRB arises from the presence of strong Kitaev correlations.  The observed BRB, its band-width, and polarization independence, are strikingly similar to the theoretically predicted Raman response of the Kitaev spin-liquid (Fig.~2 in Ref.~[23]).  The real material however, is magnetically ordered at low temperature.  We later show that the Raman response of the material in the magnetic side of the zigzag-SL boundary is qualitatively similar to predictions for the QSL.  This indicates that Na$_2$IrO$_3$ is close to the QSL state expected in the large Kitaev limit and that the BRB is a signature of persisting dynamically fluctuating moments arising from a strong competition between Heisenberg and Kitaev correlations in the material.  If we assign the BRB to be coming from Kitaev correlations and assume that the theory of Ref.~[23] stays robust for the real material, then from the peak position of the BRB we obtain a first experimental estimate of the Kitaev interaction strength of be $J_K \sim 57$~meV\@.  For magnetic ordering to survive the Heisenberg term can be expected to be $J_H \leq 0.2J_K \sim 12$~meV.  In a conventional Kitaev plus nearest (J$_H$), next nearest (J$_2$), and next next nearest (J$_3$)  neighbour Heisenberg model, the Curie Weiss temperature $\theta_{CW} = - S(S + 1)(J_H + 2J_2 + J_3 + J_K/3)$.  The observed zig zag phase is stable for a range of values of J$_2$ and J$_3$ including the values J$_2$=J$_3$= 0.5J$_H$ \cite{kimchi2011kitaev}.  For these values, and for J$_H$ = 12 meV, and with the value $J_K = 57$~meV inferred from the Raman spectrum, we find that $\theta_{CW} \approx -100$~K which is close to the experimentally observed value \cite{singh2010antiferromagnetic, singh2012relevance}.  We note that without a substantial J$_2$ and J$_3$, a small ferromagnetic value of $\theta_{CW}$ is obtained, inconsistent with experiments \cite{singh2010antiferromagnetic, singh2012relevance}.  A compelling case for large J$_K$  also  results from a similar exercise for the Chaloupka et al \cite{chaloupka2013zigzag} model which we describe in Eq. 1 of this paper, for which we have done extensive mean field calculations, and which has only nearest neighbor interactions.  Here, $\theta_{CW} = - S(S + 1)(J_H + J_K/3)$, and there is a natural parametrization  $J_H = A\cos\phi$, $J_K = 2A\sin\phi$.  For  $\theta_{CW} = -116$~K (the observed value \cite{singh2010antiferromagnetic}) and J$_K$ = 57 meV, we find $J_H = A\cos\phi = - 5.7 meV$ and $\phi$= 101.5. The latter indeed does give a zig zag phase as observed.    

We next test our hypothesis that if the BRB is a signature of Kitaev correlations, Li substituted samples should show an enhanced BRB.  It can be clearly seen from Fig.~\ref{Fig-BRB-Na2IrO3} (a) that the BRB is stronger for the Li substituted ($x = 0.1$5) crystal.  This enhancement of the BRB signal was also observed for the $x = 0.05$ crystal (not shown). The integrated  intensity (A) of the BRB (obtained by fitting a Lorentz function to the Raman susceptibility) normalized with respect to that of the $490$~cm$^{-1}$ phonon mode is shown as a function of $x$ in the inset of Fig.~\ref{Fig-BRB-T dependence}, showing an increase by $\approx 60\%$ in the $x = 0.15$ system compared to the parent $x = 0$ compound.   
The peak position of the BRB, which is given by the Kitaev interaction strength $J_K$, itself does not change noticeably with Li substitution (the BRB for $x = 0.15$ occurs at $\sim 2670$~cm$^{-1}$).  This surprisingly suggests that Li substitution tunes the relative strength of $J_K$ compared to the Heisenberg term $J_H$ without enhancing the magnitude of Kitaev term.  This can in principle be achieved if the Ir-O-Ir bond-angles reduce towards $90^o$ with Li doping but the Ir-Ir bond-lengths do not change appreciably.  This will pre-dominantly reduce the super-exchange Heisenberg term but the direct-exchange Heisenberg and Kitaev terms may not change appreciably.  Thus, both $\theta$ and $T_N$ will be affected but not the BRB position.

We now present our calculation of the Raman response for the K-H model. Our starting point is the two parameter Hamiltonian due to Chaloupka et al \cite{chaloupka2013zigzag} 
\begin{align}
H = A\cos\phi \sum {\bf{S}}_{i} \cdot {\bf{S}}_{j}  + 2A\sin\phi \sum S_{i}^a  S_{j}^a 
\label{khmodel}
\end{align}
where $S_i^a$ is the $a$ component of the spin half operator at site $i$; $i$ and $j$ are nearest neighbours.  The two parameters are the overall magnitude $A$ as well as the relative strength (and sign) of the Kitaev ($J_K = 2A\sin\phi$) and Heisenberg ($J_H = A\cos\phi$) parts of the Hamiltonian described by the angle $\phi$. The zigzag phase occurs for $\phi > 92.2^o$ using exact diagonalization \cite{chaloupka2013zigzag}. Rau et al. \cite{rau2014generic} on the other hand work on a Kitaev-Heisenberg model with off diagonal bond directional interactions, 
\begin{align}
\mathcal{H} & = \sum_{\langle ij\rangle^a} \left[ J_H  {\bf{S}}_{i} \cdot {\bf{S}}_{j}  + J_K S_{i}^a  S_{j}^a + \Gamma (S_i^b S_j^c + S_i^c S_j^b) \right]
\label{khmodel2} 
\end{align}
and find using exact diagonalization a zig zag phase, close to the spin liquid-zig zag boundary for the ground state. Though most of our discussions pertain to the model of Eq.\eqref{khmodel}, we have also done calculations with Eq.\eqref{khmodel2}.

Knolle et al \cite{knolle2014} perturbatively compute the Raman response for the spin liquid phase of Eq.\eqref{khmodel} with $J_K = -1, J_H = 0.1$ with $\phi = -78.7^o$. We use a mean field theory because of its versatility and because it gives (only) the observed phases through the actual critical values of the self consistent coupling constants separating the phases. They are different from that obtained in finite size exact diagonalization calculations.  In the Hamiltonian Eq.\eqref{khmodel}, we write the spins in terms of Majorana operators using the relations
\begin{align}
\sigma_i^\alpha &= ic_ib_i^\alpha , ~~~~~\,\{c_i,c_j\}=2\delta_{ij}, \nonumber\\
\{b^\alpha_i,b^\beta_j\}&=2\delta_{\alpha\beta}\delta_{ij}, ~~\{c_i,b^\alpha_j\}=0
\end{align}
Since the Hilbert space of states is enhanced, the physical subspace is defined by the constraint
\begin{align}
c_ib_i^xb_i^yb_i^z | \psi\rangle_{\rm phys} & =|\psi\rangle_{\rm phys}
\end{align}
The Hamiltonian in terms of the Majorana fermions is given by
\begin{align}
\mathcal{H} & = J_K \sum_{\langle ij\rangle^\alpha} ic_ib_i^\alpha ic_jb_j^\alpha + J_H \sum_{\langle ij\rangle}\sum_\alpha ic_i b_i^\alpha ic_jb_j^\alpha 
\end{align}
To solve the model we perform a mean field decoupling of the Majorana fermions to include the possibility of both spin liquid and magnetic phases as
\begin{align}
\sigma_i^\alpha\sigma_j^\beta=-ic_ic_j~ib^\alpha_ib^\beta_j
&\approx -ic_ic_jB^{\alpha\beta}_{ij}-iC_{ij}b_i^\alpha b_j^\beta+C_{ij}B^{\alpha\beta}_{ij} \nonumber \\
& + ic_ib_i^\alpha M_j^\beta + ic_j b_j^\beta M_i^\alpha - M_i^\alpha M_j^\beta
\end{align}
The self-consistency equations are
\begin{align}
& B^{\alpha\beta}_{\langle ij\rangle^\gamma}\equiv\langle ib^\alpha_ib^\beta_{j^\gamma}\rangle,
~~ C_{\langle ij\rangle^\gamma}\equiv\langle ic_ic_{j^\gamma}\rangle \nonumber \\
&  \qquad\qquad~~ M_i^\alpha \equiv \langle ic_ib_i^\alpha \rangle 
\end{align}
Here $B_{ij}$ and $C_{ij}$ represent nearest neighbour correlations and $M_i$ the magnetic order parameter.  When $M_i = 0$, the phase is a spin liquid phase where the $c$ and the $b$ Fermion Hamiltonians decouple.  The $c$ Fermions modify the hopping of the $b$ Fermions and vice versa.  When $B_{ij} = C_{ij} = 0$, the phase is a magnetic phase and the type of order is determined by the variation of $M_i$ throughout the lattice.  

In the Hamiltonian Eq.\eqref{khmodel}, we find a spin liquid to zigzag transition at $\phi = 101.4^o$. The phase transition from spin liquid to zigzag phase is first order due to the underlying symmetry of the phases.  We calculate the Raman response in the standard way \cite{knolle2014}.  

The Raman Response is given by 
\begin{align}
I(\omega) & = \int ~dt~ e^{i\omega t}~ iF(t) \\
iF(t) & = \langle GS|R(t)R(0)|GS\rangle \label{rr}
\end{align}
where $|GS\rangle$ is the ground state of the model and the Raman operator is given by\cite{knolle2014}
\begin{align}
R & =  \sum_{\langle ij \rangle \alpha} ({\boldsymbol{\epsilon}}_{in}.{\boldsymbol{d}}^\alpha) ({\boldsymbol{\epsilon}}_{out}.{\bf{d}}^\alpha) (K S_i^\alpha S_j^\alpha + K_1  {\bf{S}}_i \cdot {\bf{S}}_j)
\end{align}
where  constants $K\propto J_K$ and $K_1\propto J_H$.  The expression for the Raman response is written in the Heisenberg picture.

\begin{figure}[t]
\includegraphics[width=16cm]{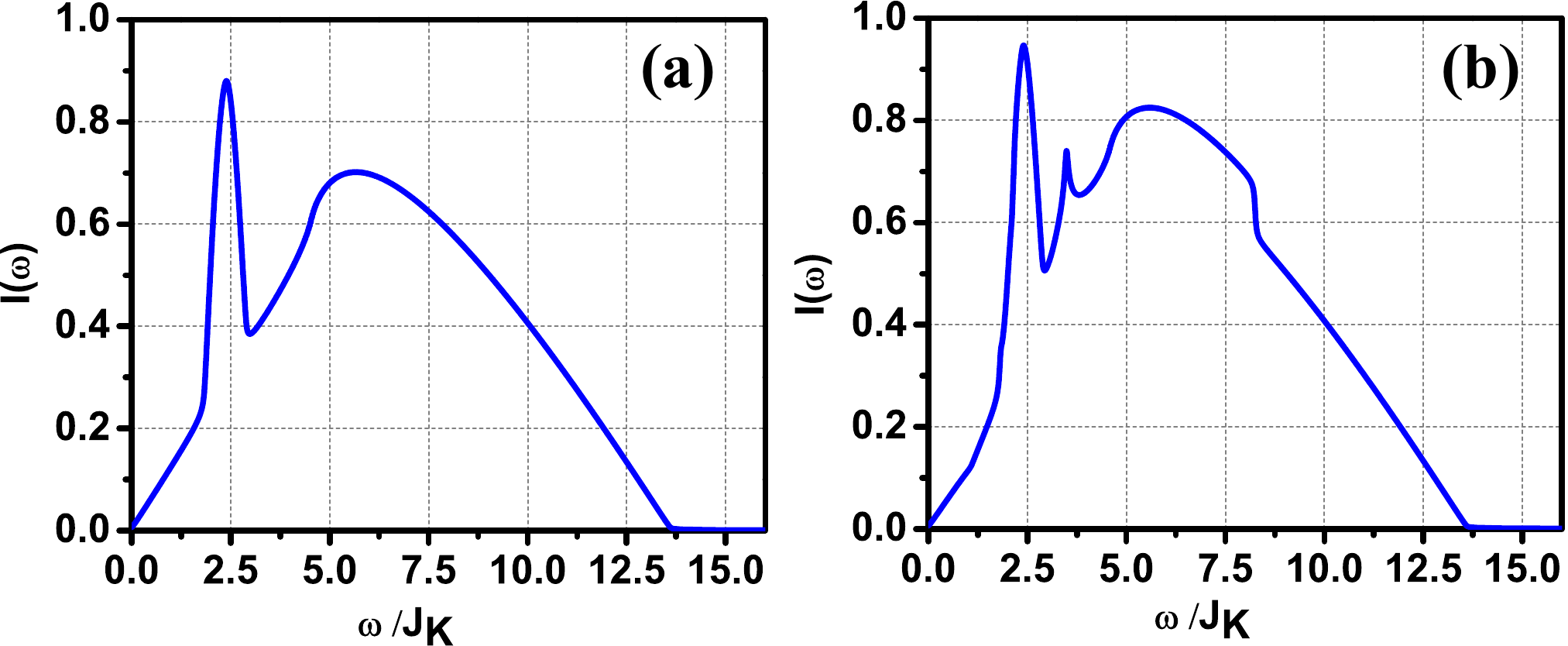}
\caption{(a) Raman Response for the Kitaev-Heisenberg model in the spin liquid phase. (b) Raman Response for the Kitaev-Heisenberg model in the zigzag phase close to spin liquid-zigzag boundary.
\label{Fig-rrsl_mixed}}
\end{figure}

In order to bring out the two striking experimental observations, namely the broad Raman spectrum and its polarization independence, we have computed the Raman spectrum in mean field theory for several polarization states.  The Raman response are shown in Fig.\ref{Fig-rrsl_mixed}.  For the SL state we find a broad polarization independent spectrum qualitatively similar to that found in Ref.~[23].  Fig.\ref{Fig-rrsl_mixed}~(b) shows the Raman spectrum for $A=1$ and $\phi = 101.5^o$, values which give zigzag phase close to spin liquid-zigzag boundary.  We note that the broad Raman mode seen for the SL, survives in the magnetic state as well.  This explains the observation of the BRB below T$_N$ in our experiments.  It is worth mentioning that the experimental BRB looks more like the Raman response for the SL state than the Raman response for the magnetic state which shows some additional structure around $3.75~ \omega/J_K$ and $9~ \omega/J_K$.  We have also found that the Raman response does not change much even if an off diagonal bond directional term ($\Gamma \approx 0.01$) is added like in Eq.\eqref{khmodel2}.  Thus the BRB survives in the magnetic state and in the presence of other small terms apart from the Kitaev term.

In summary, we have shown the existence of a broad, polarization independent Raman band at high energies for single crystals of (Na$_{1-x}$Li$_x$)$_2$IrO$_3$ ($x = 0, 0.05, 0.15$).  This observation is in excellent agreement with predictions for the observation of such a band as a signature of the Kitaev spin-liquid state in the honeycomb lattice iridates~ \cite{knolle2014}.  Similar observations have recently been made on another candidate Kitaev material $\alpha$-RuCl$_3$.  The observation of the BRB in $\alpha$-RuCl$_3$ was interpreted as resulting from proximity to the QSL phase in the strong Kitaev limit \cite{sandilands2015scattering}.  However, the real materials (both the iridates and the above ruthenate) are magnetically ordered at low temperatures.  Thus it was unclear whether this broad continuum, predicted for the SL state, would survive in the magnetically ordered state.  We have shown using mean field calculations of the K-H model that the BRB survives in the magnetically ordered state at least near the zigzag-SL phase boundary where the Na$_2$IrO$_3$ material is most likely situated \cite{chun2015direct}.  The BRB predicted for the magnetic state acquires more structure compared to the BRB in the SL state.  Our observed BRB resembles that predicted for the SL state more than it does for the magnetic phase.  This suggests that Na$_2$IrO$_3$ is proximate to the zigzag-QSL phase boundary and strong Kitaev correlations are present.  From the position of the peak of the band, we make a first direct experimental estimate of the Kitaev interaction strength to be $J_K = 57$~meV\@.  This estimate although much larger than the values estimated before in the literature ($J_K \sim -2~{\rm to} -17$~meV) [for example \cite{singh2012relevance, chaloupka2013zigzag, choi2012spin, katukuri2014kitaev} is consistent with the experimentally observed Weiss temperature.  The fact that we observe the BRB well into the magnetically ordered state is consistent with recent diffuse RIXS observations which indicate that dynamical fluctuations, present above $T_N$, survive almost unchanged into the magnetic phase \cite{chun2015direct}.  Finally, the enhanced signature in Li substituted samples suggests that these materials maybe better avenues to search for further proof for dominant Kitaev physics.
\acknowledgments
A.K.S. and T.V.R. acknowledges funding from DST. YS acknowledges partial support from DST through the Ramanujan fellowship and through the grant no. SB/S2/CMP-001/2013. S.N.G. thanks CSIR for the SRF.



\end{document}